\documentclass[tinyml]{acmart}
\usepackage{subfig}

\AtBeginDocument{%
  \providecommand\BibTeX{{%
    \normalfont B\kern-0.5em{\scshape i\kern-0.25em b}\kern-0.8em\TeX}}}

\setcopyright{rightsretained}
\copyrightyear{2023}
\acmYear{2023}

\begin{document}

\title{How Tiny Can Analog Filterbank Features Be Made \\ for Ultra-low-power On-device Keyword Spotting?}


\author{Subhajit Ray}
\affiliation{%
  \institution{Columbia University}
  \city{New York}
  \country{United States}}
\email{subhajit.ray@columbia.edu}

\author{Xinghua Sun}
\affiliation{%
  \institution{Columbia University}
  \city{New York}
  \country{United States}}
\email{xs2445@columbia.edu}

\author{Nolan Tremelling}
\affiliation{%
  \institution{Columbia University}
  \city{New York}
  \country{United States}}
\email{nnt2109@columbia.edu}

\author{Maria Gordiyenko}
\affiliation{%
  \institution{Barnard College}
  \city{New York}
  \country{United States}}
\email{mg4366@barnard.edu}

\author{Peter Kinget}
\affiliation{%
  \institution{Columbia University}
  \city{New York}
  \country{United States}}
\email{pk171@columbia.edu}


\begin{abstract}
Analog feature extraction is a power-efficient and re-emerging signal processing paradigm for implementing the front-end feature extractor in on-device keyword-spotting systems. Despite its power-efficiency and re-emergence, there is little consensus on what values the architectural parameters of its critical block, the analog filterbank, should be set to, even though they strongly influence power consumption. Towards building consensus and approaching fundamental power consumption limits, we find via simulation that through careful selection of its architectural parameters, the power of a typical state-of-the-art analog filterbank could be reduced by 33.6$\times$, while sacrificing only 1.8\% in downstream 10-word keyword spotting accuracy through a back-end neural network.
\end{abstract}


\maketitle

\section{Introduction}

Reducing the power consumption of on-device keyword spotting (KWS) to ultra-low power levels will be instrumental in realizing the promise of pervasive intelligence. KWS systems typically consist of a front-end feature extractor and back-end neural network. Despite extensive software research on neural networks for keyword spotting \cite{14ICASSP_ChenHeigold,15INTERSPEECH_SainathParada,18ICASSP_TangLin}, hardware research, specifically chip design research, reveals that a state-of-the-art KWS chip \cite{20JSSC_GiraldoVerhelst} suffers from the power bottleneck being not the back-end neural network, but rather the front-end feature extractor, consuming approximately $70$\% of total KWS system power. Therefore, research in power-efficient front-end feature extraction is important.

\begin{figure}[ht]
  \centering
  \includegraphics[width=\linewidth]{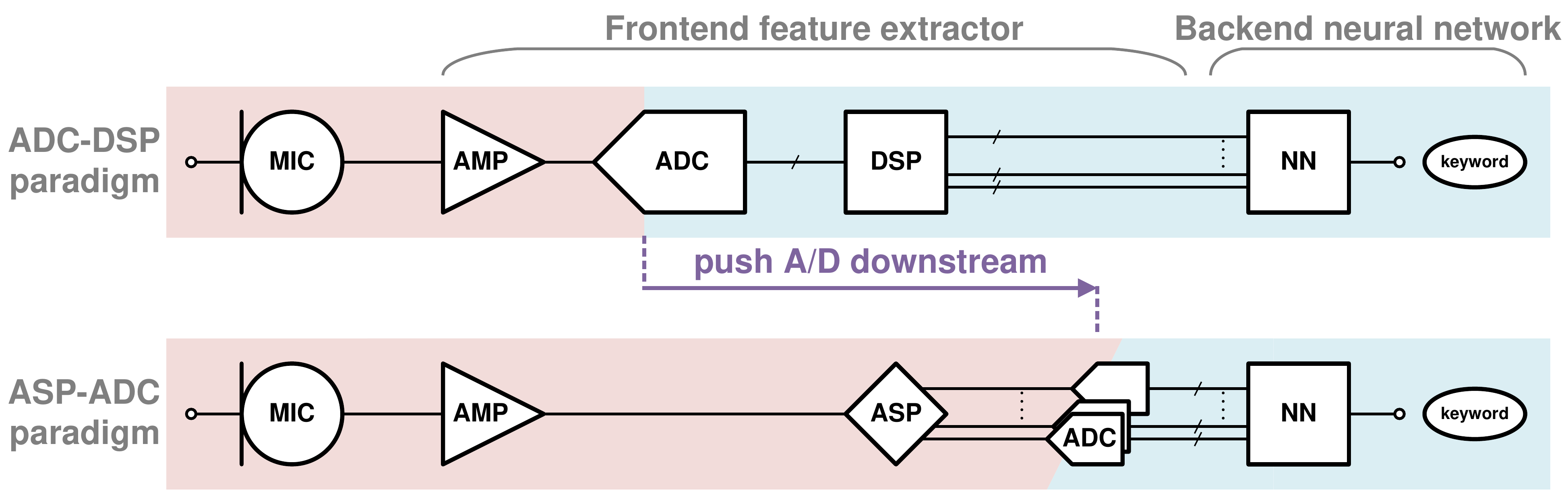}
  \caption{The conventional paradigm for front-end feature extraction using Analog-to-Digital Conversion then Digital Signal Processing (ADC-DSP) versus the unconventional but re-emerging and more power-efficient paradigm using Analog Signal Processing then Analog-to-Digital Conversion (ASP-ADC). The paradigm of interest in this paper is ASP-ADC, the critical block of which is the analog filterbank inside the ASP.}
  \label{fig:intro}
\end{figure}
More power-efficient than the conventional, digital-intensive paradigm for front-end feature extraction adopted by \cite{20JSSC_GiraldoVerhelst}, which uses Nyquist-Rate-Analog-to-Digital Conversion plus Digital Signal Processing (ADC-DSP), is the less conventional but rapidly re-emerging \cite{22JSSC_RayKinget,22JSSC_KimLiu,21JSSC_YangSeokEnz,21TCASI_VillamizarMurmann,22SSCL_Fuketa,21TCASI_ShiPun,19JSSC_YangSeok,15JSSC_BadamiVerhelst} analog-intensive paradigm, which uses Analog Signal Processing plus Feature-Rate-Analog-to-Digital Conversion (ASP-ADC) \cite{15SSCSMAG_VerhelstBahai,05TBME_Sarpeshkar}, as shown in Fig. \ref{fig:intro}. Case in point, the lowest-power ASP-ADC-based front-end feature extractor \cite{21JSSC_YangSeokEnz} is $12\times$ more power-efficient than the lowest-power ADC-DSP-based front-end feature extractor (amplifier and ADC \cite{18VLSI_BadamiVerhelst}, DSP \cite{22JSSC_ShanCai}). Although, in general, ASP-ADC does suffer from less programmability and more variability than ADC-DSP, research in ASP-ADC-based front-end feature extraction is important because of its power efficiency, and because of its broad applicability, not only to tinyML audio, but also to tinyML vision \cite{22TINYML_LuMurmann}.

The critical block in an ASP-ADC-based front-end feature extractor is the analog filterbank, which is inside the ``ASP" block in Fig. \ref{fig:intro}. Even though the filterbank's architectural parameters strongly influence power consumption, there is little consensus on what values they should be set to, as revealed by Table \ref{tab:parameters}, which shows the parameter value choices made by state-of-the-art analog feature extractor chips \cite{22JSSC_RayKinget,22JSSC_KimLiu,21JSSC_YangSeokEnz,21TCASI_VillamizarMurmann}. The lack of consensus is further surprising because all these chips target the same task of KWS. Motivated by both this lack of consensus and also the pursuit of fundamental power consumption limits, we ask how ``tiny" the analog filterbank can be made without appreciably degrading classification accuracy; and, we find that through careful selection of its architectural parameters, it could be made $33.6\times$ lower power while sacrificing only $1.8$\% in KWS accuracy.
\begin{table}
  \caption{A summary of the choices made by state-of-the-art analog feature extractor chips for the values of the main architectural parameters of the analog filterbank, revealing a lack of consensus although all chips target the same task of keyword spotting.}
  \label{tab:parameters}
  \begin{tabular}{l|ccc}
     & $N_\mathrm{filters}$ & $f_\mathrm{max,filterbank}$ & $Q_\mathrm{filter}$ \\
    \hline
    JSSC'22 \cite{22JSSC_RayKinget}        & 31 & 4kHz  & 3  \\
    JSSC'22 \cite{22JSSC_KimLiu}         & 16 & 10kHz & 2  \\
    JSSC'21 \cite{21JSSC_YangSeokEnz}        & 16 & 5kHz  & 2  \\
    TCASI'21\cite{21TCASI_VillamizarMurmann} & 32 & 8kHz  & 26 \\
    \hline
    Typical (average)          & 24 & 7kHz  & 8  \\
\end{tabular}
\end{table}

The remainder of the paper is organized as follows: section \ref{sec:background} provides background on analog feature extraction and filterbanks; section \ref{sec:simulation_methodology} discusses the simulation methodology; section \ref{sec:simulation_results} reports the simulation results and proposes a tiny analog filterbank and compares it to a typical one; section \ref{sec:discussion} dicusses the results in a broader context; finally, section \ref{sec:conclusion} recapitulates the key points of the paper.

\section{Background}\label{sec:background}

\subsection{Analog Feature Extraction}
In the ASP-ADC paradigm for front-end feature extraction, colloquially referred to as analog feature extraction, the block labeled ``ASP" in Fig. \ref{fig:intro} is generally composed of an analog filterbank followed by an analog envelope detector bank. Consisting of a set of bandpass filters, the function of the filterbank is to decompose the input signal into its frequency components. Subsequently, the function of the envelope detector bank is to detect the envelope, a measure of average power, of each frequency component. As such, the analog feature extractor output is a feature vector, where each element has a value proportional to the short-term average power as a function of time inside a frequency sub-band. And so, by stacking feature (column) vectors laterally over time, what results is a spectrogram of the input signal, which serves as the input to the back-end neural network.

Each of the two constituent blocks of an analog feature extractor, the analog filterbank and analog envelope detector bank, has its own architectural parameters. The architectural parameters of the envelope detector, its window length and window overlap, already see consensus among state-of-the-art analog feature extractor chips, with typical values being 20-32ms and 10-16ms, respectively \cite{22JSSC_RayKinget,22JSSC_KimLiu,21TCASI_VillamizarMurmann}. Not only because of this consensus, but also because these parameters only weakly influence the power of the envelope detector, we choose not to explore its parameters.

\subsection{Analog Filterbanks}\label{sec:background_analog_filterbanks}

\begin{figure*}[h]
\centering
\subfloat[$N_\mathrm{filters}=64$\label{fig:filterbanks_high_n_filter}]{\includegraphics[width=0.3\linewidth]{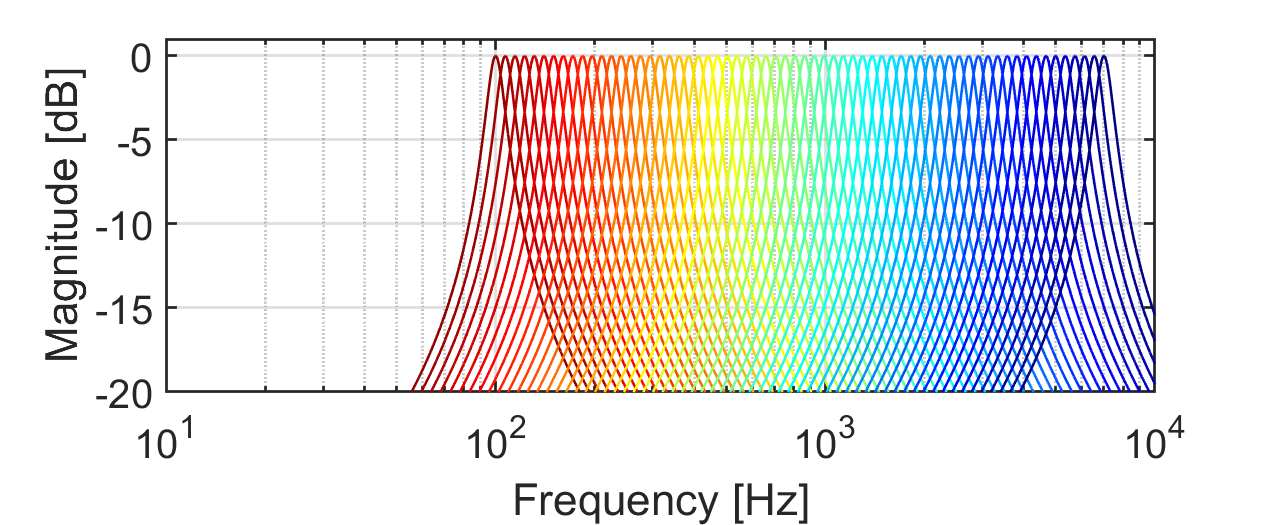}}\hfil
\subfloat[$f_\mathrm{max,filterbank}=10$kHz\label{fig:filterbanks_high_f_max_filterbank}]{\includegraphics[width=0.3\linewidth]{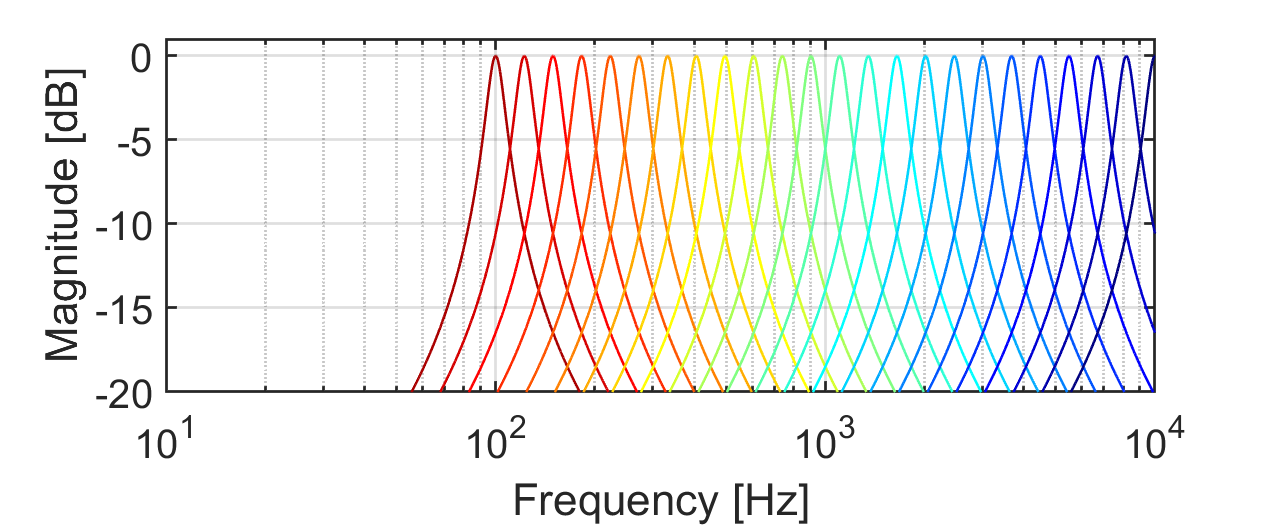}}\hfil 
\subfloat[$Q_\mathrm{filter}=60$\label{fig:filterbanks_high_q_filter}]{\includegraphics[width=0.3\linewidth]{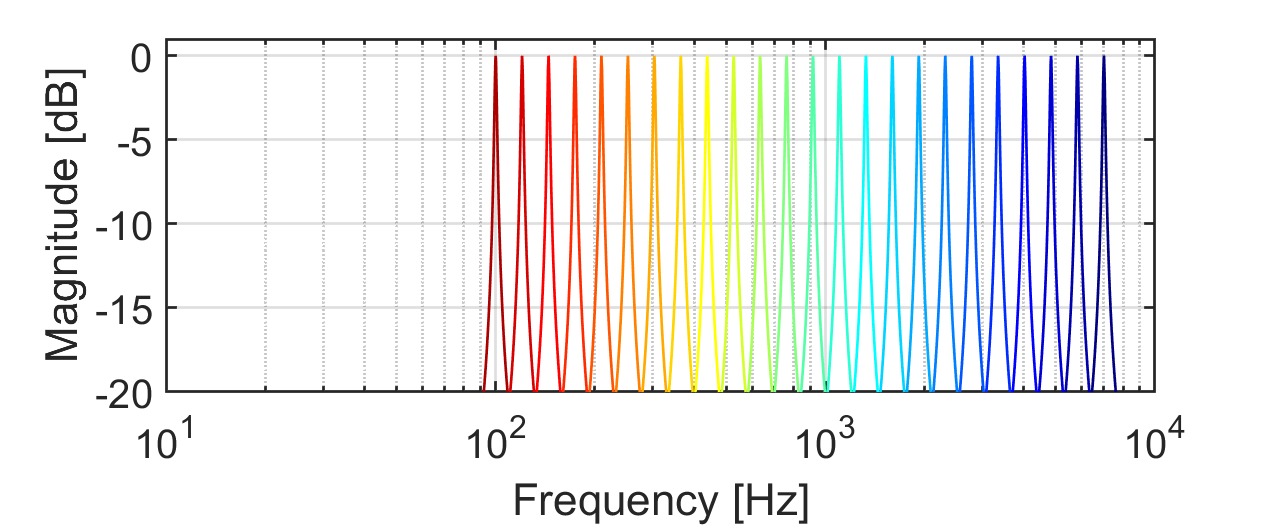}}\\
\subfloat[$N_\mathrm{filters}=4$\label{fig:filterbanks_low_n_filter}]{\includegraphics[width=0.3\linewidth]{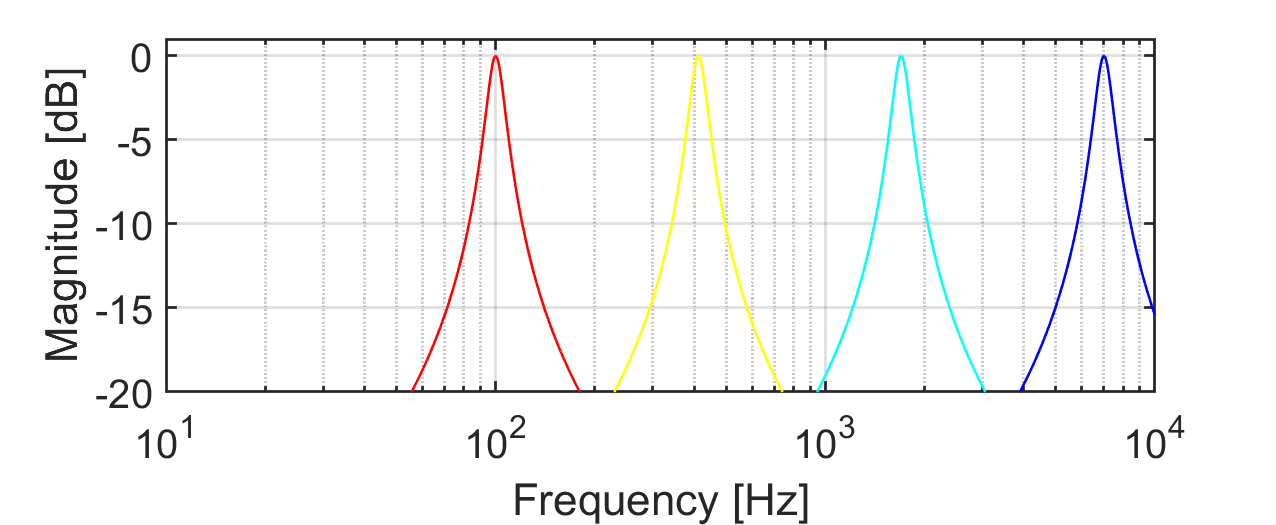}}\hfil   
\subfloat[$f_\mathrm{max,filterbank}=0.25$kHz\label{fig:filterbanks_low_f_max_filterbank}]{\includegraphics[width=0.3\linewidth]{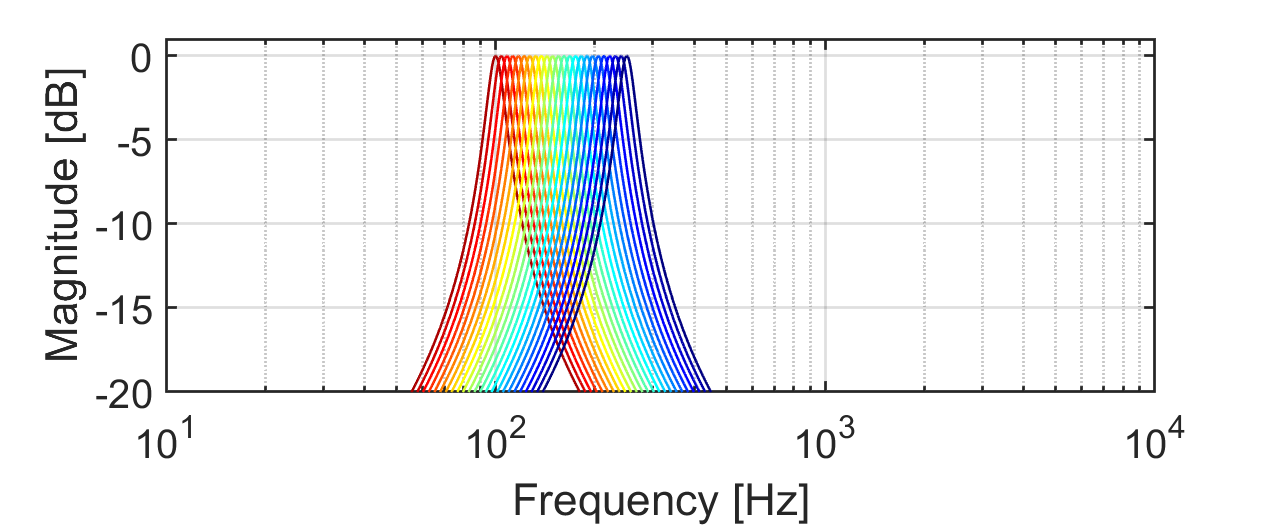}}\hfil
\subfloat[$Q_\mathrm{filter}=0.2$\label{fig:filterbanks_low_q_filter}]{\includegraphics[width=0.33\linewidth]{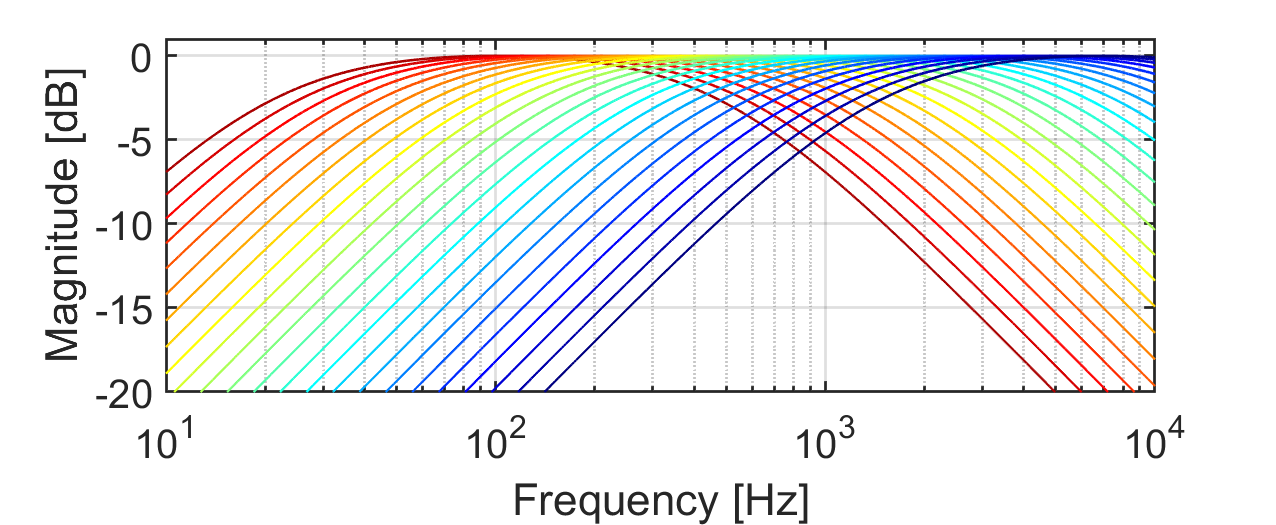}}
\caption{Frequency responses (ideal) of an analog filterbank across extreme values of its architectural parameters, to illustrate the meaning of each parameter. Filterbank power consumption is proportional, to first order, to each of these parameters.}
\label{fig:filterbanks}
\end{figure*}

On the other hand, there is little consensus on what values the architectural parameters of the filterbank should be set to, as was shown in Table \ref{tab:parameters}. This lack of consensus is important to tinyML because the power consumption of a filterbank is proportional\footnote{Provided each filter's SNR is kept constant.}, to first order, to each of the three\footnote{Provided the other two parameters are kept constant.} architectural parameters of interest \cite{18SSCSMAG_Tsividis}: $N_\mathrm{filters}$, $f_\mathrm{max,filterbank}$, $Q_\mathrm{filter}$. We draw an important distinction between these \textit{architectural} parameters and those that capture the non-idealities of the analog filterbank, namely noise and distortion, which also strongly influence power consumption. Because in the analog design process, architectural parameters are typically specified first before specifying requirements for noise and distortion, we choose in this paper to first explore the architectural parameters, leaving noise and distortion for future research.

To better understand the architectural parameters of interest, Fig. \ref{fig:filterbanks} depicts the frequency responses of a filterbank across extreme parameter values. Note that in the caption of each sub-figure, when a parameter value is left unstated, then its value is set to its typical value specified in Table \ref{tab:parameters}. The parameter $N_\mathrm{filters}$ is the number of bandpass filters that comprise the filterbank. The parameter $f_\mathrm{max,filterbank}$ is the maximum frequency of the filterbank, which, more specifically, is the center frequency of the highest-frequency bandpass filter in the filterbank. The parameter $Q_\mathrm{filter}$ is the quality factor of all bandpass filters in the bank. Quality factor is a measure of the frequency selectivity of a bandpass filter, and is defined as the ratio of its center frequency to its 3dB-bandwidth, which is to say quality factor is the reciprocal of the fractional bandwidth.

Besides the three architectural parameters of interest, there are a few secondary ones; however, they already see consensus on what values to set them to or have a weak influence on filterbank power consumption. The secondary architectural parameters, along with their typical value ranges are given by the following:
\begin{itemize}
\item The center frequency of the lowest-frequency bandpass filter. It is commonly set to 30-100Hz in state-of-the-art analog feature extractor chips \cite{22JSSC_RayKinget,22JSSC_KimLiu,21JSSC_YangSeokEnz,21TCASI_VillamizarMurmann}. Its influence on filterbank power is weak because the lowest-frequency bandpass filter's power is a small fraction of the total power of all bandpass filters in the bank.
\item The distribution of the center frequencies. It is commonly set to an exponential distribution\footnote{This is why in Fig. \ref{fig:filterbanks} the center frequencies appear linearly distributed on the log frequency scale.} in state-of-the-art analog feature extractor chips \cite{22JSSC_RayKinget,21JSSC_YangSeokEnz}, as motivated by the biological cochlea \cite{98SIGPROCMAG_loizou}. However, some do use Mel \cite{22JSSC_KimLiu} or ERB \cite{21TCASI_VillamizarMurmann} distributions.
\item The order of the bandpass filter. It is commonly set to second order in state-of-the-art analog feature extractor chips \cite{22JSSC_RayKinget,22JSSC_KimLiu,21JSSC_YangSeokEnz}. This is because such chips already attain adequate KWS accuracy, while higher orders would consume proportionally more power.
\item The quality factor distribution. It is common to set the quality factors of all bandpass filters in the bank to the same value\footnote{This is why in Fig. \ref{fig:filterbanks} the 3dB bandwidths of all the bandpass filters in the bank appear to be the same on the log frequency axis.}, in state-of-art-art analog feature extractor chips \cite{22JSSC_RayKinget,22JSSC_KimLiu,21JSSC_YangSeokEnz}, which is motivated by the biological cochlea \cite{98SIGPROCMAG_loizou}.
\end{itemize}

Having discussed each of the three analog filterbank feature parameters of interest, $N_\mathrm{filters}$, $f_\mathrm{max,filterbank}$, $Q_\mathrm{filter}$, and recognized that filterbank power consumption is proportional, to first order, to each, the question now is, to what extent can each parameter value be reduced without appreciably degrading KWS accuracy through a back-end neural network. We answer this question by sweeping each of the three parameters, while holding the other two constant at their typical values specified in Table \ref{tab:parameters}, and simulating KWS accuracy for each sweep point, in order to reveal the parameter values below which accuracy degrades appreciably.

\section{Simulation Methodology}\label{sec:simulation_methodology}

\subsection{Pre-Feature Extraction Dataset}\label{sec:simulation_methodology_dataset}
The pre-feature extraction dataset of audio signals was constructed from the Google Speech Commands Dataset (GSCD) v0.02, which consists of 1-second-long, 16-bit, 16kS/s audio signals \cite{18arXiv_Warden}. Of its 35 available words, 10 of them---yes, no, up, down, left, right, on, off, stop, go---were designated as keyword classes, with the remaining 25 designated as the 11th unknown class. For each keyword class, 200 training examples were used, while for the unknown class, more training examples, 800 of them, were used; otherwise, the classification accuracy for the unknown class would suffer relative to the those for the keyword classes, which is so because many (25) different words comprise the unknown class. Taken together, the pre-feature extraction training set consists of 2,800 examples.

Though the GSCD offers at least 3,000 examples per word, the reason only 200 were used was to limit feature extraction and training times, as constrained by the available computing resources. With so many examples left over, we chose to additionally use 200 examples per keyword in the pre-feature extraction test (and validation) set, instead of, for example, 20. Resulting in a training:validation:test set ratio of 1:1:1, this choice was made to make use of the leftover examples to improve the estimates of the test set accuracies shown in Figs. \ref{fig:acc_vs_n_filters}, \ref{fig:acc_vs_f_max_filterbank}, \ref{fig:acc_vs_q_filter}.

Finally, although the dataset underlying Figs. \ref{fig:acc_vs_n_filters}, \ref{fig:acc_vs_f_max_filterbank}, \ref{fig:acc_vs_q_filter} was kept relatively small at 2,800 training examples to speed up the required 47 training sessions, a 10$\times$-larger dataset with 28,000 training examples was used for the final comparison in Figs. \ref{fig:typical_vs_tiny_power_and_accuracy} and \ref{fig:typical_vs_tiny_more_detail} between the ``typical" analog filterbank and the proposed ``tiny" one.

\subsection{Front-end Analog Feature Extractor}
The pre-feature extraction dataset (training/validation/test) was fed through our custom MATLAB model of an analog feature extractor to produce a post-feature extraction dataset of spectrograms on which the back-end neural network was trained and tested. As described in section \ref{sec:background}, the analog feature extractor model consists of a programmable filterbank followed by a fixed envelope detector bank. Further, as was explained in section \ref{sec:background_analog_filterbanks}, the noise and distortion of the filterbank are not modeled and left for future research; in this way, the effect of the filterbank's architectural parameters on KWS accuracy is isolated.

The fixed parameters of the envelope detector are: 20ms window length, 10ms window overlap. The fixed parameters of the filterbank are: 100Hz center frequency of the lowest-frequency bandpass filter, exponential distribution of the center frequencies of all the bandpass filters in bank, 2nd-order bandpass filters, and all bandpass filters in the bank have the same quality factor.

The programmable parameters of the filterbank are $N_\mathrm{filters}$, \newline$f_\mathrm{max,filterbank}$, and $Q_\mathrm{filter}$. Each of the three was swept over a range of values, while the other two were set to their typical values specified by Table \ref{tab:parameters}. The ranges of values were: \newline$N_\mathrm{filters}=\{1,2,4,6,8,10,12,14,16,20,24, 28,32,48,64\}$, $f_\mathrm{max,filterbank} = \{0.25,0.5,1,2,3,4,5,6,7,8,9,10,12,16,20\}\mathrm{kHz}$, and \newline$Q_\mathrm{filter} = \{0.2,0.4,0.6,0.8,1,2,3,4,6,8,10,15,20,25,30,40,60\}$. These  ranges were selected such that they extend well below and above the values chosen by the state-of-the-art analog feature chips, shown in Table \ref{tab:parameters}.

Each of the 47 filterbank parameter sweep points corresponds to a different instance of the analog feature extractor model, and therefore a different post-feature extraction dataset, generated by processing the pre-feature extraction dataset with the feature extractor instance.

\subsection{Back-end Neural Network}

The back-end neural network is re-trained on each of the 47 post-feature extraction training sets, and subsequently tested on the 47 post-feature extraction test sets. Moreover, the neural network is re-trained and re-tested across three trials per sweep point to construct a confidence interval for the KWS accuracy at each point. We choose to do this because analog feature extractor chip papers published so far have not been reporting KWS accuracies with confidence intervals \cite{22JSSC_RayKinget,22JSSC_KimLiu,21JSSC_YangSeokEnz,21TCASI_VillamizarMurmann}.

To isolate the effect of the filterbank from the effect of the back-end neural network, the same neural network architecture was used for all filterbank parameter sweep points. The architecture used was the small-footprint residual convolutional neural network \texttt{res16} of \cite{18ICASSP_TangLin}, specifically our implementation of it using the MATLAB Deep Learning Toolbox \cite{matlab_deep_learning_toolbox}. This architecture was selected because it was used by the most power-and-area-efficient published analog audio feature extractor chip in its KWS demonstration.

To isolate the effect of the filterbank from the effect of the training algorithm hyperparameters, the same hyperparameter values were used for all filterbank parameter sweep points. The hyperparameter values used were those obtained from manual tuning  on the post-feature extraction validation set produced by the feature extractor model with the typical filterbank parameter values ($N_\mathrm{filters}=24$, $f_\mathrm{max,filterbank}=7\mathrm{kHz}$, $Q_\mathrm{filter}=8$) plugged in. The hyperparameter values that resulted from the tuning were: momentum of $0.9$, learning rate with an initial value of $1$ that drops by a factor of $0.9$ after each epoch, $25$ epochs, minibatch size of $64$, $L_2$ regularization of $0.001$, and the training algorithm used was stochastic gradient descent with momentum.

\section{Simulation Results}\label{sec:simulation_results}
Figs. \ref{fig:acc_vs_n_filters}, \ref{fig:acc_vs_f_max_filterbank}, \ref{fig:acc_vs_q_filter} show the intermediary simulation results: plots of 10-word KWS accuracy vs each of the three analog filterbank architectural parameters of interest. The purpose of these plots is to guide how tiny an analog filterbank can be made without appreciably degrading classification accuracy. Fig. \ref{fig:typical_vs_tiny_power_and_accuracy} shows the final simulation result of the paper, which is a comparison between the typical state-of-the-art filterbank and the proposed tiny filterbank in terms of relative power and downstream accuracy through a back-end neural network

As section \ref{sec:simulation_methodology_dataset} explained, and we re-iterate here, the final simulation result of Fig. \ref{fig:typical_vs_tiny_power_and_accuracy} used the larger 28,000-example training set, whereas the intermediary simulation results of Figs. \ref{fig:acc_vs_n_filters}, \ref{fig:acc_vs_f_max_filterbank}, \ref{fig:acc_vs_q_filter} used the smaller 2,800-example training set, which was done to speed up the 141 training sessions required to produce the data for those plots. Because of this training set size difference, and because neural networks favor more training data, the typical-filterbank-accuracy of 94.2\% seen in Fig. \ref{fig:typical_vs_tiny_power_and_accuracy} is 3.3\% higher than the 90.9\% accuracy seen across Figs. \ref{fig:acc_vs_n_filters}, \ref{fig:acc_vs_f_max_filterbank}, \ref{fig:acc_vs_q_filter} at the red lines, even though both accuracy numbers correspond to the same filterbank, the typical one.

\subsection{Number of filters}\label{sec:results_n_filters}
\begin{figure}[h]
  \centering
  \includegraphics[width=\linewidth]{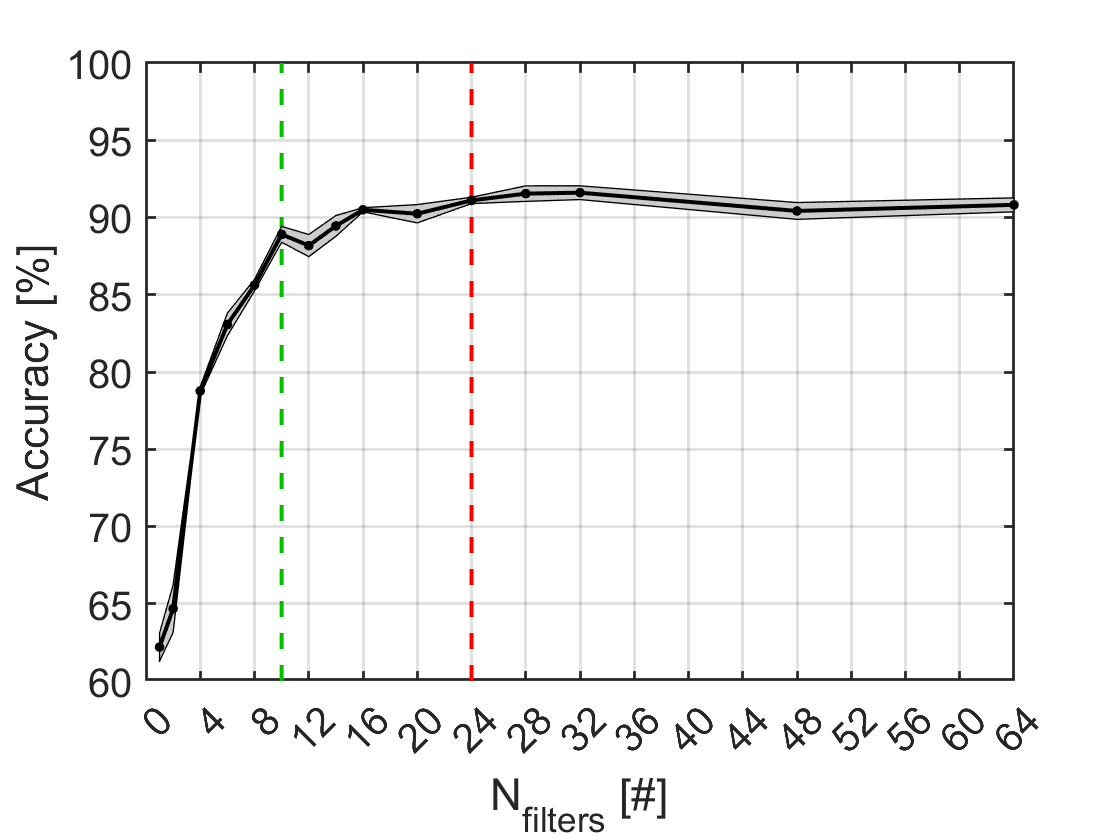}
  \caption{10-word KWS accuracy vs $N_\mathrm{filters}$. Results are based on the 2,800-example training set. The 3-trial-based 95\% confidence interval bounds are shown. The typical choice made in state-of-the-art analog feature extractor chips is indicated in red, and the proposed tiny choice in green.}
  \label{fig:acc_vs_n_filters}
\end{figure}
Fig. \ref{fig:acc_vs_n_filters} shows the plot of 10-word KWS classification accuracy vs the number of filters in the filterbank, $N_\mathrm{filters}$. As expected, the accuracy is lower for fewer filters and higher for more, as fewer and more frequency sub-bands are processed, respectively, as can be visualized in Figs. \ref{fig:filterbanks_low_n_filter} and \ref{fig:filterbanks_high_n_filter}. Although it's expected based on heuristic reasoning that more features would exhibit diminishing marginal returns in accuracy, it is surprising that there is a plateau such that the accuracy for as many as 64 filters is approximately the same as that for as few as 16 filters. This suggests that it would be power-wasteful for 10-word KWS to have a filterbank with $N_\mathrm{filters}$ more than 16. Finally, it's interesting to note that even for just a single feature, the accuracy is 62.2\%, whereas, for context, random spotting of keywords would give 9.1\% accuracy (11 classes, including the 10 keywords, and the unknown class).

\subsection{Maximum frequency of filterbank} \label{sec:results_f_max_filterbank}
\begin{figure}[h]
  \centering
  \includegraphics[width=\linewidth]{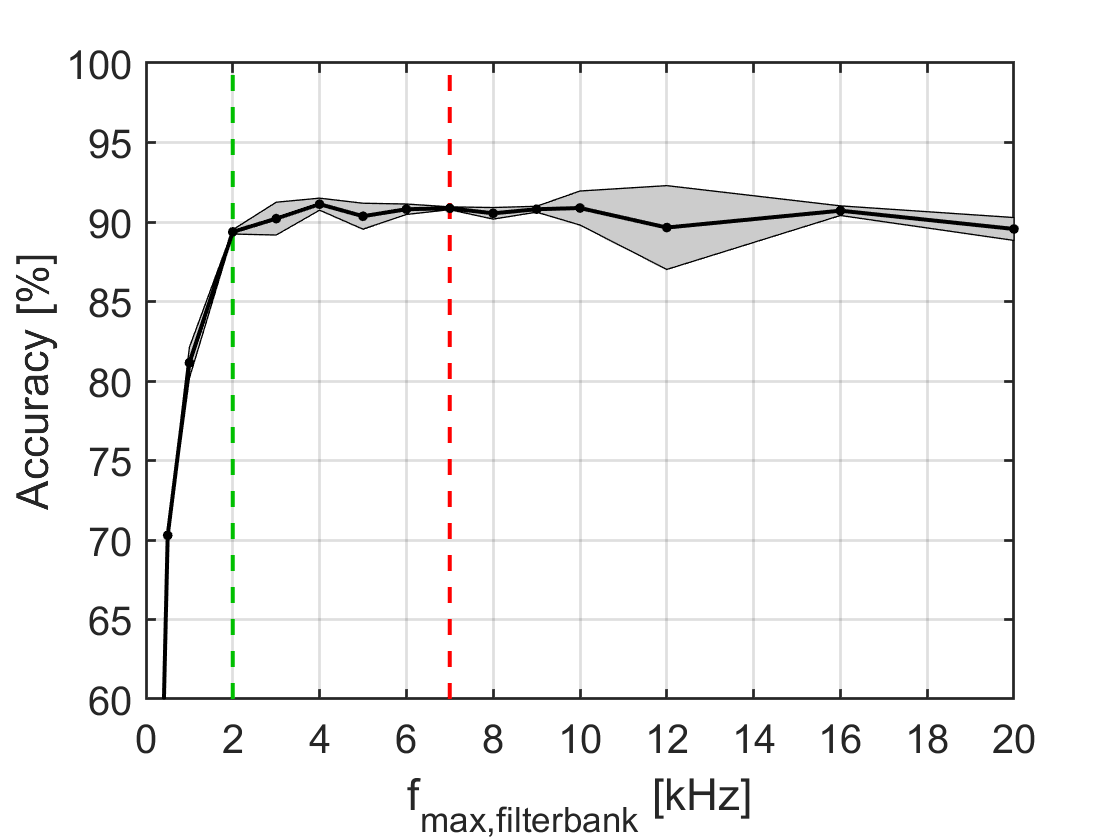}
  \caption{10-word KWS accuracy vs $f_\mathrm{max,filterbank}$. Results are based on the 2,800-example training set. The 3-trial-based 95\% confidence interval bounds are shown. The typical choice made in state-of-the-art analog feature extractor chips is indicated in red, and the proposed tiny choice in green.}
  \label{fig:acc_vs_f_max_filterbank}
\end{figure}
Fig. \ref{fig:acc_vs_f_max_filterbank} shows the plot of 10-word KWS classification accuracy vs the maximum frequency of the filterbank, $f_\mathrm{max,filterbank}$. As expected, the accuracy is lower for lower $f_\mathrm{max,filterbank}$ and higher for higher $f_\mathrm{max,filterbank}$, as narrower and wider fractions of the speech bandwidth would be processed, respectively, as can be visualized in Figs. \ref{fig:filterbanks_low_f_max_filterbank} and \ref{fig:filterbanks_high_f_max_filterbank}. In contrast to the $N_\mathrm{filters}$ plot (Fig. \ref{fig:acc_vs_n_filters}) where the existence of the plateau is surprising, in this  $f_\mathrm{max,filterbank}$ case it's not, as speech bandwidth is limited to $4\mathrm{kHz}$ \cite{98SIGPROCMAG_loizou}, meaning that if $f_\mathrm{max,filterbank}$ is made excessively high, then the highest frequency filter outputs will be zero-valued, conveying zero information, and therefore not increase accuracy. However, what is surprising given that the speech bandwidth is $4\mathrm{kHz}$ is that the accuracy doesn't appreciably degrade until below only $2\mathrm{kHz}$. This suggests that it would be power-wasteful for 10-word KWS to have a filterbank with an $f_\mathrm{max,filterbank}$ greater than $2\mathrm{kHz}$.

\subsection{Quality factor of filter}\label{sec:results_q_filter}
\begin{figure}[h]
  \centering
  \includegraphics[width=\linewidth]{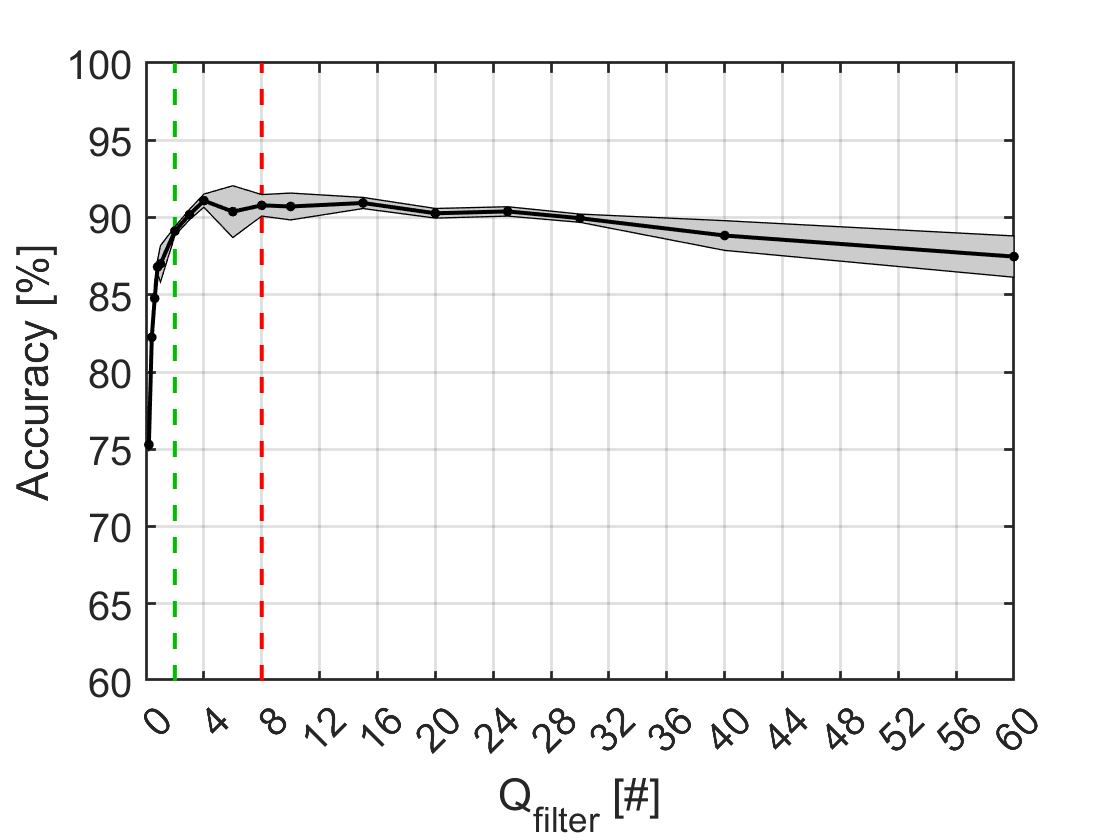}
  \caption{10-word KWS accuracy vs $Q_\mathrm{filter}$. Results are based on the 2,800-example training set. The 3-trial-based 95\% confidence interval bounds are shown. The typical choice made in state-of-the-art analog feature extractor chips is indicated in red, and the proposed tiny choice in green.}
  \label{fig:acc_vs_q_filter}
\end{figure}
Fig. \ref{fig:acc_vs_q_filter} shows the plot of 10-word KWS classification accuracy vs the quality factor of the filters in the filterbank, $Q_\mathrm{filter}$. As expected, the accuracy is lower for lower $Q_\mathrm{filter}$ because the filters are less selective and become redundant with one another, as can be visualized in Fig. \ref{fig:filterbanks_low_q_filter}. Now, observe that for excessively high $Q_\mathrm{filter}$ the accuracy drops. This, too, is expected because for excessively high $Q_\mathrm{filter}$, the filterbank's passbands are extremely narrow, and consequently the inter-passband frequency gaps are extremely wide, as can be visualized in Fig. \ref{fig:filterbanks_high_q_filter}, and in these gaps the signal energy is lost. Thus, there should exist an optimal $Q_\mathrm{filter}$, and Fig. \ref{fig:acc_vs_q_filter} shows that it is $4$, below which the accuracy degrades rapidly	 (and above which it degrades shallowly). This suggests that it would be power-wasteful for 10-word KWS to have a filterbank with $Q_\mathrm{filter}$ more than $4$.

\subsection{Typical vs tiny filterbank}

\begin{figure}
  \centering
  \includegraphics[width=\linewidth]{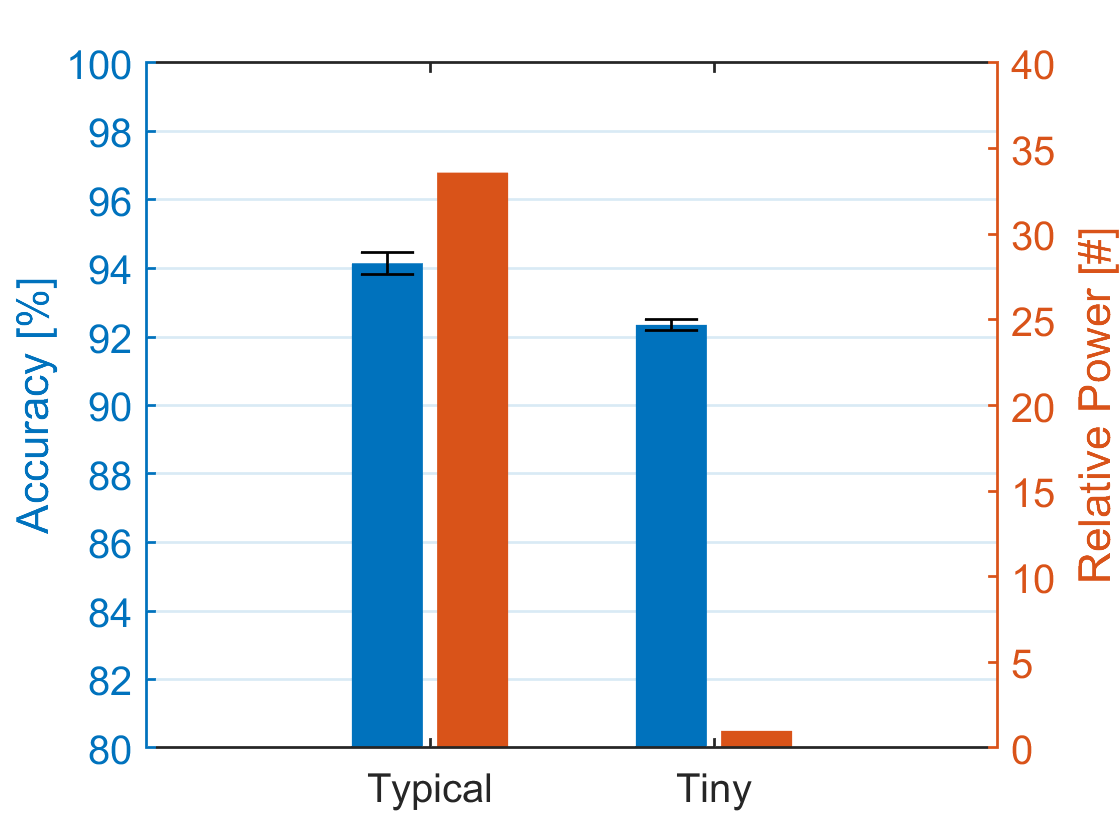}
  \caption{Comparison between a typical state-of-the-art analog filterbank and the proposed tiny one in terms of relative power and downstream 10-word KWS accuracy through a back-end neural network. Results are based on the larger 28,000-example dataset. The 5-trial-based 95\% confidence intervals are annotated.}
  \label{fig:typical_vs_tiny_power_and_accuracy}
\end{figure}

\begin{figure*}[h]
\centering
\subfloat[Frequency response of typical filterbank]{\includegraphics[width=0.3\linewidth]{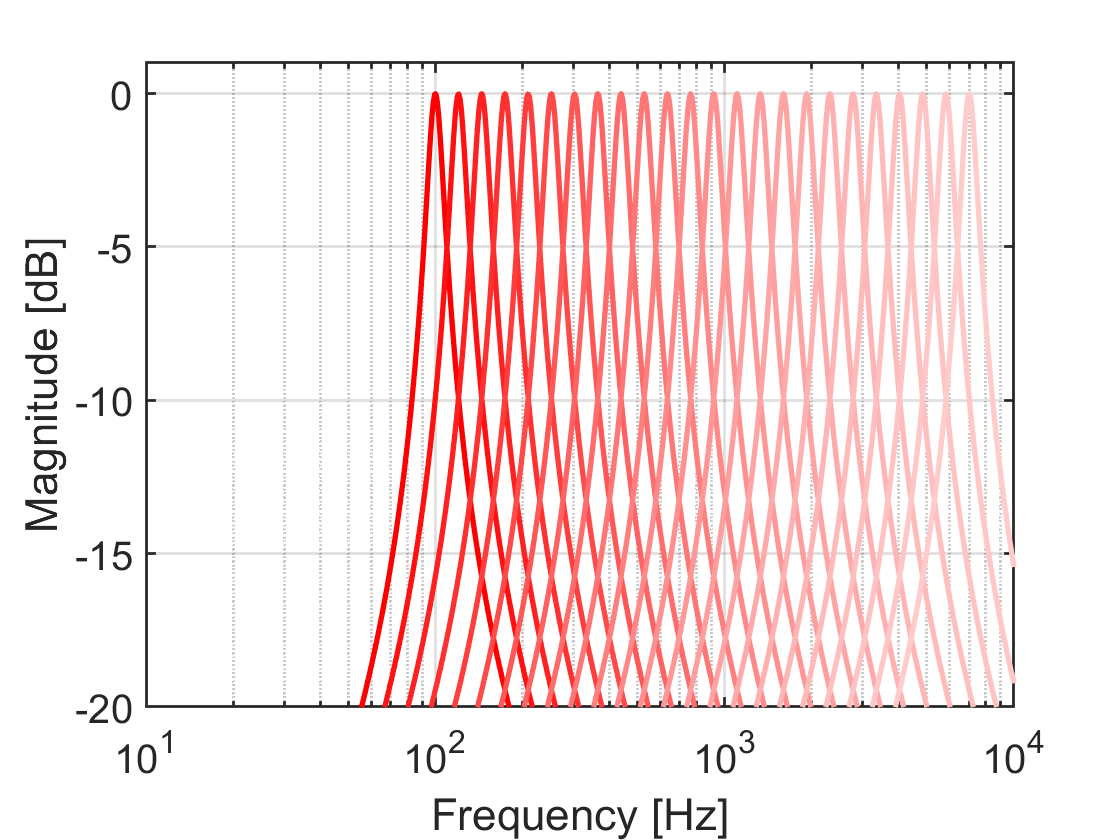}}\hfil
\subfloat[Spectrogram of ``yes" using typical filterbank]{\includegraphics[width=0.3\linewidth]{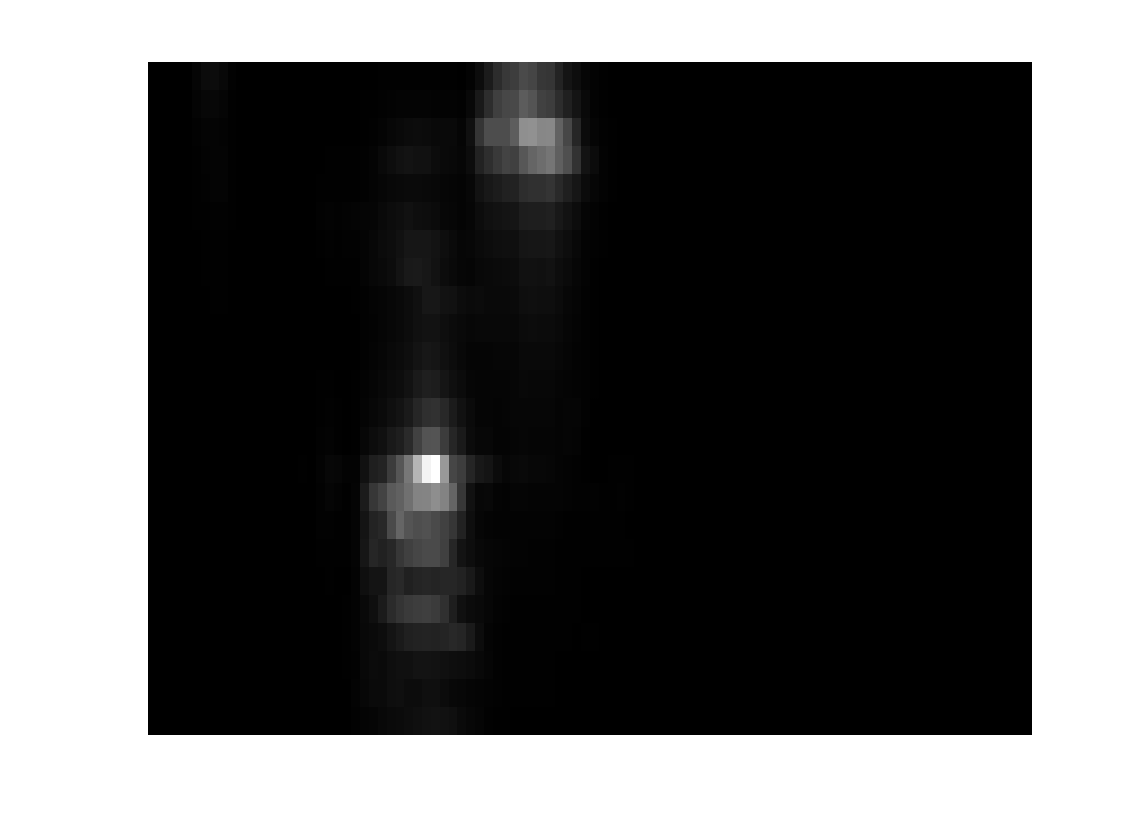}}\hfil 
\subfloat[Confusion matrix using typical filterbank]{\includegraphics[width=0.3\linewidth]{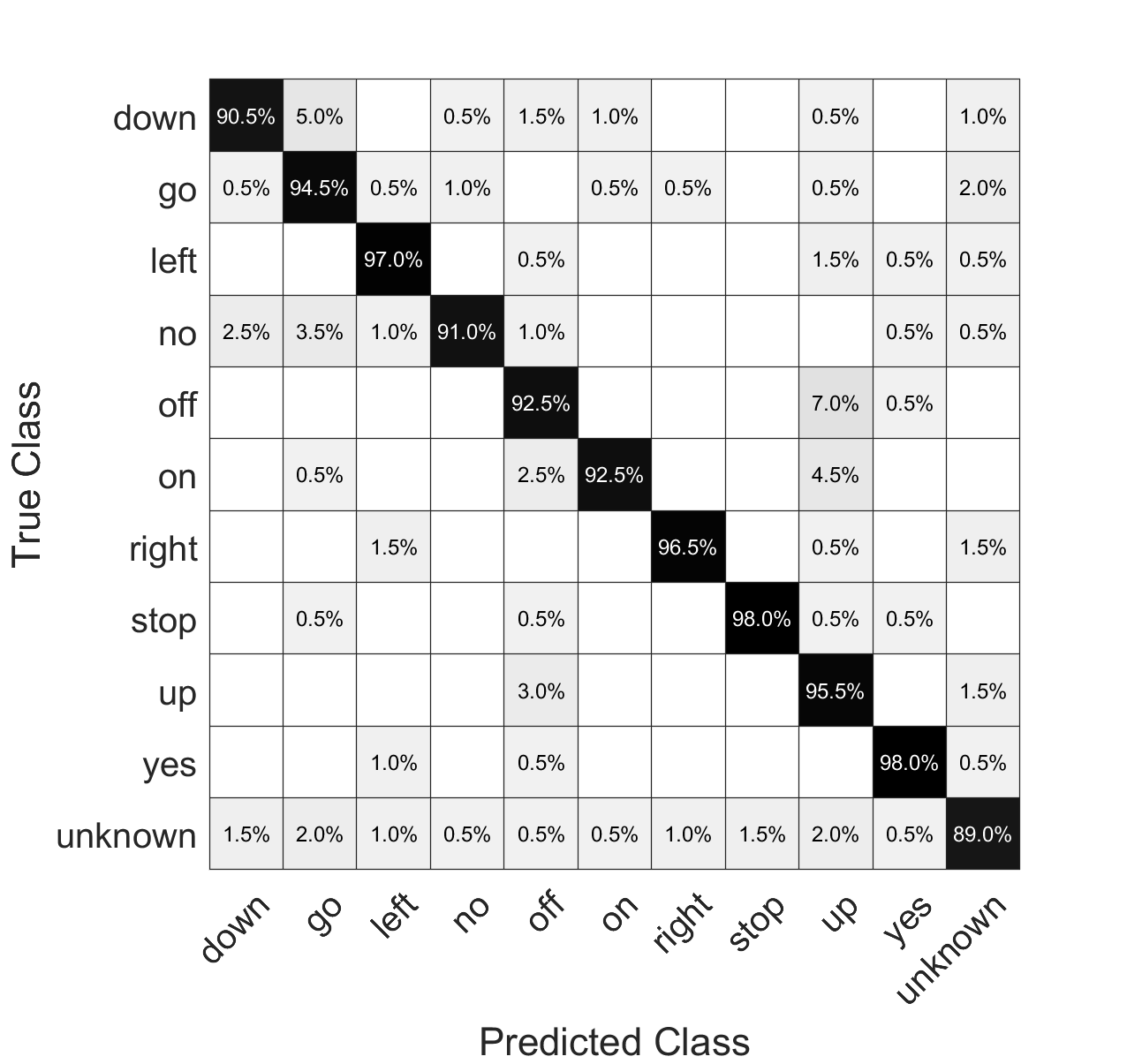}}\\
\subfloat[Frequency response of tiny filterbank]{\includegraphics[width=0.3\linewidth]{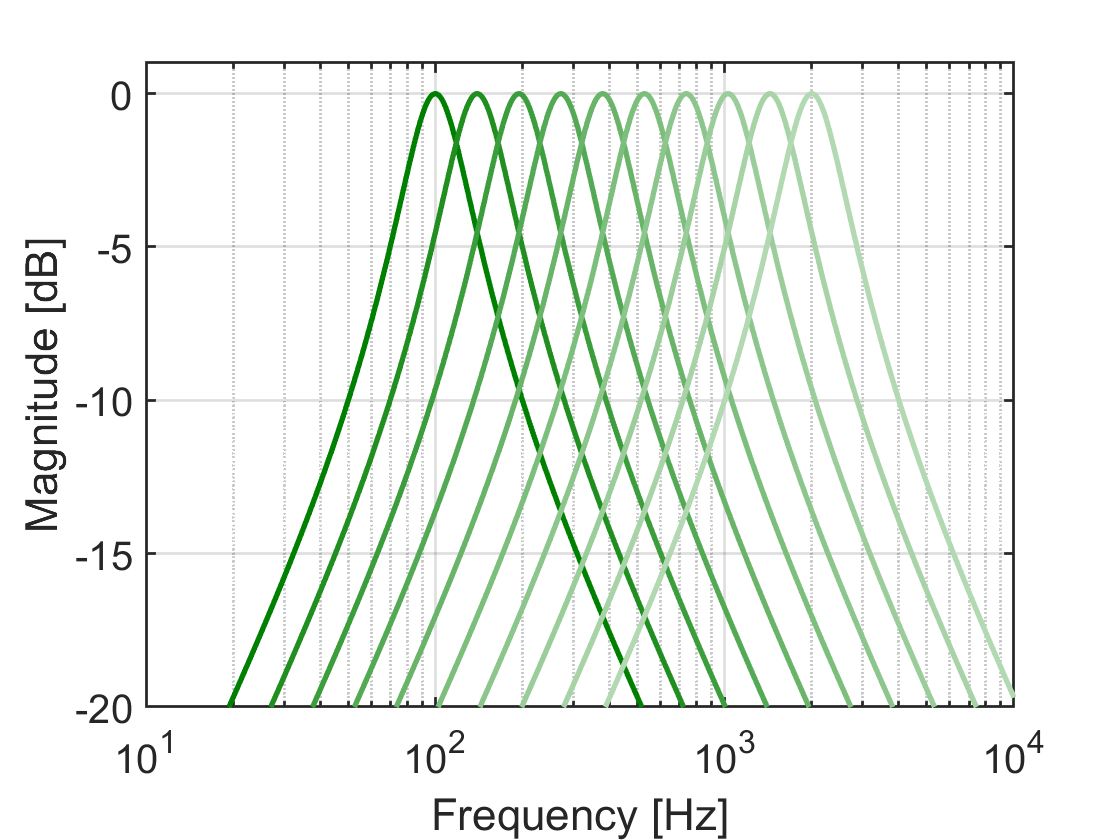}}\hfil   
\subfloat[Spectrogram of ``yes" using tiny filterbank]{\includegraphics[width=0.3\linewidth]{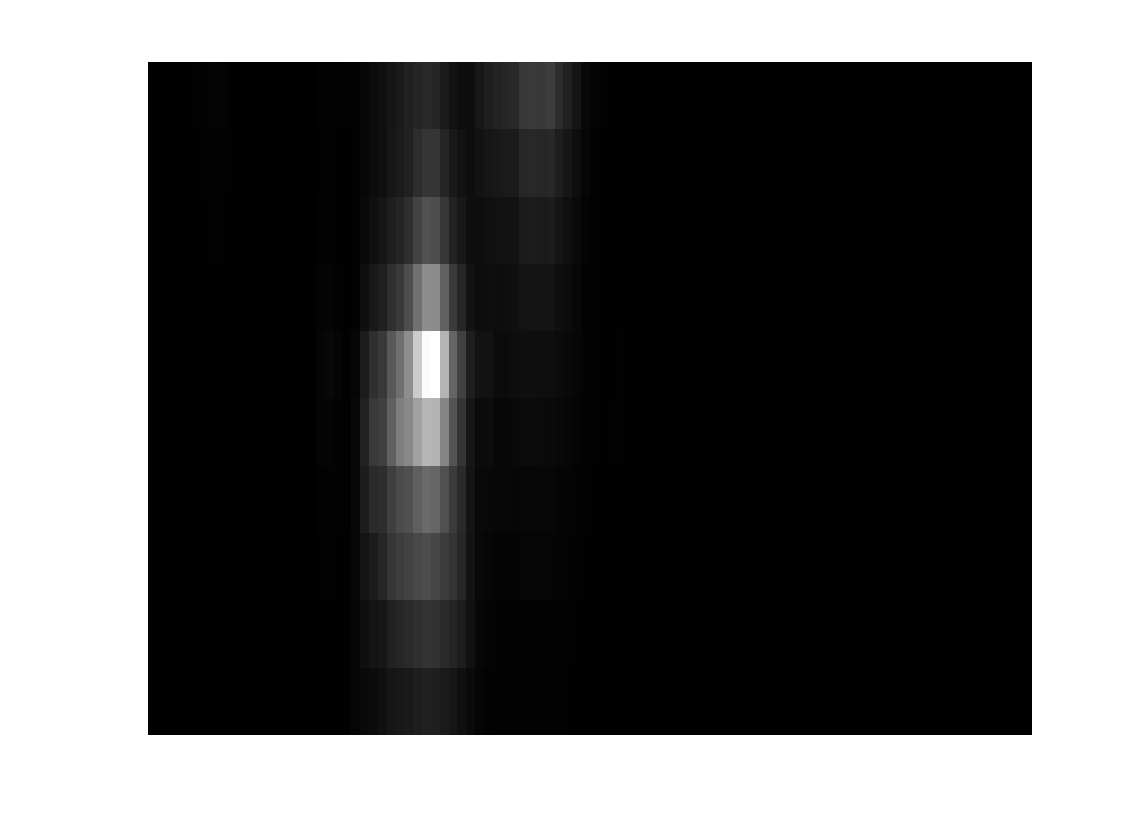}}\hfil
\subfloat[Confusion matrix using tiny filterbank]{\includegraphics[width=0.33\linewidth]{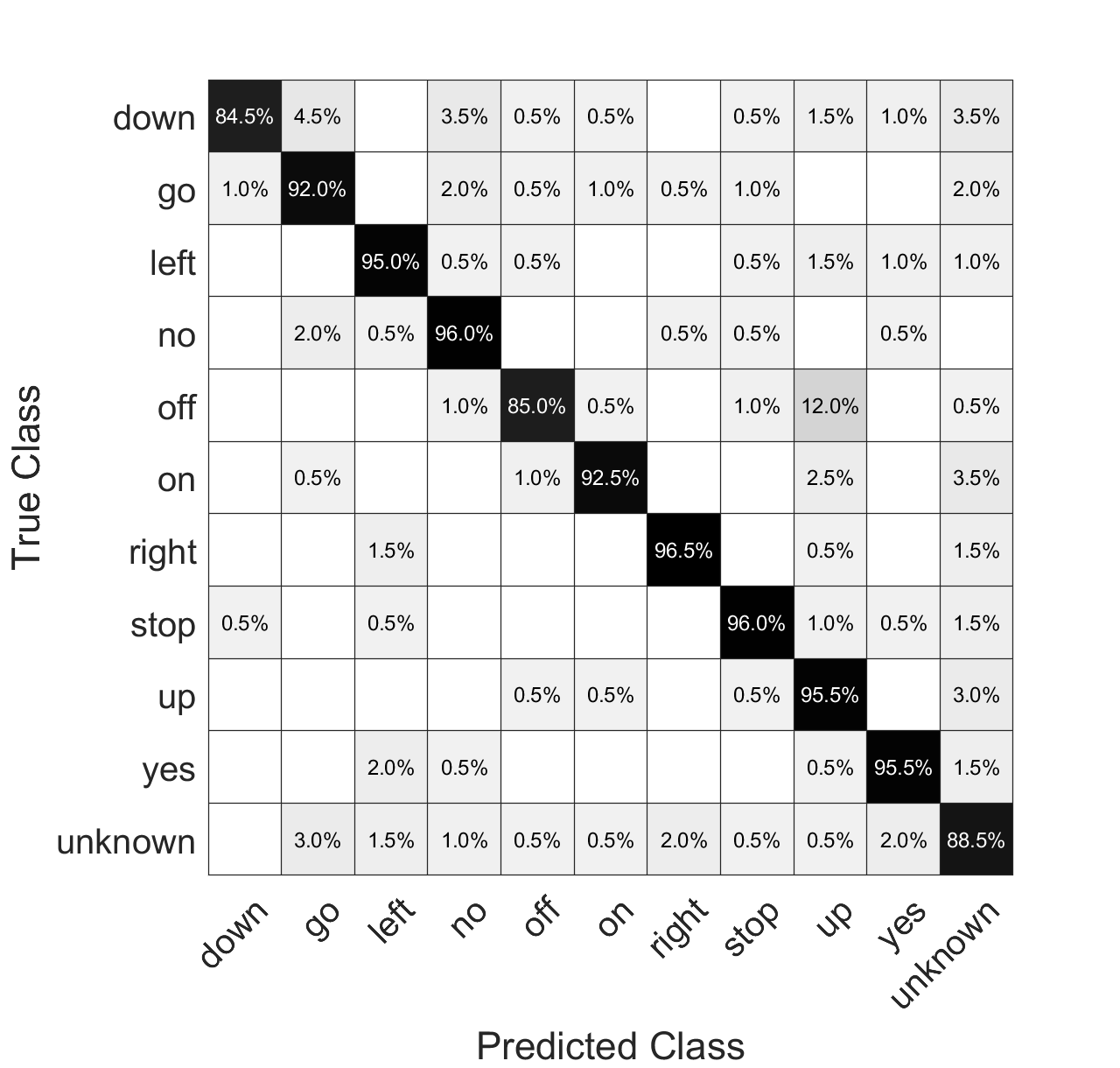}}
\caption{Comparison between a typical state-of-the-art analog filterbank and the proposed tiny one in terms of frequency response (a)(d), an example spectrogram with time on the x axis and frequency on the y axis (b)(e), and confusion matrix using a back-end neural network whose inputs are the spectrograms (c)(f).}
\label{fig:typical_vs_tiny_more_detail}
\end{figure*}

To summarize, Figs. \ref{fig:acc_vs_n_filters}, \ref{fig:acc_vs_f_max_filterbank}, \ref{fig:acc_vs_q_filter} show that the typical values of the analog filterbank architectural parameter values, indicated by the red lines, can be reduced significantly to those indicated by the green lines, so as to define the resulting ``tiny" filterbank with $N_\mathrm{filters}=10$, $f_\mathrm{max,filterbank}=2\mathrm{kHz}$, and $Q_\mathrm{filter}=2$. Because these tiny values correspond to $2.4\times$, $3.5\times$, and $4\times$ reductions, respectively, relative to the typical values, the tiny filterbank should be $33.6\times$ lower power than the typical filterbank, recalling from section \ref{sec:background_analog_filterbanks} that filterbank power consumption is proportional, to first order, to each of the three parameters.

Fig. \ref{fig:typical_vs_tiny_power_and_accuracy} shows that the $33.6\times$-lower-power tiny filterbank degrades downstream 10-word KWS accuracy through the back-end neural network by only 1.8\% relative to the typical filterbank, which is a modest price to pay for such extreme power savings. Fig \ref{fig:typical_vs_tiny_more_detail} goes on to compare the typical and tiny filterbanks in terms of frequency response (a)(d), spectrogram (b)(e), and confusion matrix (c)(f). The frequency responses visually depict that that of the tiny filterbank (d) has a smaller number of filters, a lower maximum frequency, and lower quality factor than that of the typical filterbank (a). This leads to the tiny filterbank spectrogram (e) being ``lower resolution" than the typical spectrogram (b). Even then, the two distinct regions---corresponding to the lower-frequency ``ye.." and the higher-frequency ``...s" parts of ``yes"---that are obvious to the human eye in the typical spectrogram (b), can still be made out in the tiny spectrogram (e)\footnote{Though, the higher-frequency ``...s" region is admittedly tough to discern.}. This exercise may impart some intuition on why the tiny filterbank doesn't significantly degrade accuracy. Finally, for completeness, (c) and (f) show the confusion matrices corresponding to the typical and tiny filterbanks, respectively.

\section{Discussion}\label{sec:discussion}
The $33.6\times$ filterbank power reduction is large while the $1.8$\% accuracy degradation is small, suggesting that this is a good trade-off to make. However, it would ultimately be the application requirements that would determine whether a $33.6\times$ filterbank power reduction is worth the $1.8$\% accuracy degradation. Further, it should be stipulated that the results of this paper are specific to the chosen back-end neural network, which is the small-footprint residual convolutional neural network \texttt{res15} of \cite{18ICASSP_TangLin} published 4 years ago. Although the results could change for a different neural network, say a more modern one, it is likely that the essential conclusion would remain the same, that a large power reduction can be won at the cost of a small accuracy degradation. Finally, we note some interesting tangential research: \cite{21arXiv_BergsmaCernak} analyzes  filterbank features in terms of information theory, and \cite{21EUSIPCO_Lopez-EspejoJensen} compares handcrafted filterbank features to learned filterbank features.

\section{Conclusion}\label{sec:conclusion}
In the context of KWS systems with a front-end analog feature extractor and back-end digital neural network, we find through simulation that by carefully selecting the values of its architectural parameters, the power consumption of the typical analog filterbank in state-of-the-art analog feature extractor chips can be reduced by $33.6\times$, while sacrificing only 1.8\% in downstream 10-word KWS accuracy through the neural network. 

Having explored the effect of filterbank architectural parameters on KWS accuracy, this work sets the stage to next explore the effect of the filterbank non-idealities, as measured by SNR and linearity, on KWS accuracy. Additionally, this work raises a host of new questions; for example, how would the results change for a higher number of keywords, say 30; or, how would they change for other audio recognition tasks like voice activity detection, wake word detection, or speaker identification.  Finally, although this work focuses on and optimizes the front-end feature extractor, it paves the way for co-optimization with the back-end neural network.

\begin{acks}
Rui Xu is thanked for simulation support, and Rebecca Zhang is thanked for software neural network implementation support. This research was supported in part by NSF 1704899 and Analog Devices. 
\end{acks}

\bibliographystyle{ACM-Reference-Format}
\bibliography{refs}

\end{document}